\documentclass[draftcls, onecolumn]{IEEEtran}

\usepackage[pdftex]{graphicx}
\DeclareGraphicsExtensions{.pdf,.jpeg,.png}

\usepackage[cmex10]{amsmath}
\usepackage{amsfonts}
\usepackage{cite}
\usepackage{url}
\usepackage{color}
\graphicspath{{figures/}}

\newcommand{\BE}{\begin{equation}}
\newcommand{\EE}{\end{equation}}
\newcommand{\BF}{\begin{figure}\centering}
\newcommand{\EF}{\end{figure}}
\newcommand{\BT}{\begin{table}\centering}
\newcommand{\ET}{\end{table}}

\newcommand{\D}[1]{\, \mathrm{d} #1} 
\newcommand{\MAT}[1]{\mathbf{#1}} 
\newcommand{\MR}[1]{\mathrm{#1}} 
\newcommand{\J}{\jmath} 

\newcommand{\TM}{$\mathrm{TM}_{10}$ }     
\newcommand{\TE}{$\mathrm{TE}_{10}$ }     
\newcommand{\WM}{W_\mathrm{m}}     
\newcommand{\WE}{W_\mathrm{e}}     
\newcommand{\Wrad}{W_\mathrm{rad}} 
\newcommand{\Wom}{W_{\omega}}      
\newcommand{\PM}{P_\mathrm{m}}     
\newcommand{\PE}{P_\mathrm{e}}     
\newcommand{\Prad}{P_\mathrm{rad}} 
\newcommand{\Pom}{P_{\omega}}      
\newcommand{\Q}{Q}                 
\newcommand{\Qhansen}{Q_\mathrm{HC}} 
\newcommand{\Qcollin}{Q_\mathrm{Chu}} 

\newcommand{\QRc}{Q_R} 
\newcommand{\QXc}{Q_X} 
\newcommand{\QXcU}{Q_X^\mathrm{untuned}}   
\newcommand{\QXcT}{Q_X^\mathrm{tuned}} 
\newcommand{\QZc}{Q_Z} 
\newcommand{\QZcU}{Q_Z^\mathrm{untuned}} 
\newcommand{\QZcT}{Q_Z^\mathrm{tuned}} 

\newcommand{\Cx}[1]{\cos\left( #1 x \right)}
\newcommand{\Sx}[1]{\sin\left( #1 x \right)}

\author{
    \IEEEauthorblockN{Lukas~Jelinek, Miloslav~Capek, Pavel~Hazdra, Jan~Eichler \\}
    \IEEEauthorblockA{Department of Electromagnetic Field, Faculty of Electrical Engineering, Czech Technical University in Prague, Technicka 2, 16627, Prague, Czech Republic \\ Email: lukas.jelinek@fel.cvut.cz}
}

\title{An Analytical Evaluation of The Quality Factor $Q_Z$ for Dominant Spherical Modes}

\begin{document}

\maketitle

\begin{abstract}
This paper describes an analytical evaluation of the quality factor $Q_Z$ in a separable system in which the vector potential is known. The proposed method uses a potential definition of active and reactive power, implicitly avoiding infinite entire space integration and extraction of radiation energy. As a result, all the used quantities are finite, and the calculated $Q_Z$ is always non-negative function of frequency. The theory is presented on the canonical example of the currents flowing on a spherical shell. The $Q_Z$ for the dominant spherical TM and TE mode and their linear combination are found in closed forms, including both internal and external energies. The proposed analytical method and its results are compared to previously published limits of the quality factor $Q$.
\end{abstract}

\section{Introduction}
\label{Intro}
The quality factor ($\Q$ factor) is recognized as one of the most significant parameters of a radiating system, especially if the electrical dimensions are small, see e.g. \cite{VolakisChenFujimoto_SmallAntennas_MiniatrurizTechniques} and references therein. The reason is its approximate inverse proportionality to a fractional bandwidth (FBW) and a possibility of establishing the lower bounds of $\Q$ factor \cite{VolakisChenFujimoto_SmallAntennas_MiniatrurizTechniques}. This implies an upper bound of FBW, a restriction of substantial importance for electrically small antennas (ESAs).

The classical work on lower bounds of $\Q$ factor is the work of Chu \cite{Chu_PhysicalLimitationsOfOmniDirectAntennas}, which considers a sphere of radius $a$ that encloses an ESA. The normalized radial wave impedance for the dominant spherical TM mode is expressed as a continued fraction equivalent to a ladder network with particular $R$,\,$L$,\,$C$ elements. In this way, the  lower bound of $\Q$  can be found. However, the Chu's method is restricted to the spherical modes only and does not include the internal energy of the sphere, making the limit overly optimistic. Later, Wheeler \cite{Wheeler_TheRadiansphereAroundASmallAntenna} reduced the basic radiators, dipole and loop, to the circuit elements and derived practically oriented limits. Expansion to the spherical harmonics was also used by Harrington \cite{Harrington_EffectsOfAntennaSizeOnGainBWandEfficiency} to evaluate the electric and magnetic energy for each mode. The same approach was presented by Collin and Rothschild \cite{CollinRotchild_EvaluationOfAntennaQ} for spherical and cylindrical modes. McLean \cite{McLean_AReExaminationOfTheFundamentalLimitsOnTheRadiationQofESA} verified the Chu's formula. He obtained the same results, but his approach is based on the field radiated by the Herzian dipole. Thiele, Detweiler and Penno \cite{ThieleDetweilerPenno_OnTheLowerBoundOfTheRadiationQforESA} used the ``far-field method", based on the separation of the far-field pattern into its visible and invisible parts \cite{Rhodes_OnTheStoredEnergyOfPlanarApertures}. Thal \cite{Thal_RadiationQandGainOfTMandTEsourcesInPhaseDelayedRotatedConfigurations}, \cite{Thal_CommentsOnNewChuFormula}, \cite{Thal_QboundsForArbitrarySmallAntennas}  used the ladder network to  extend the Chu's  limit by including the energies inside the  enclosing  sphere. Hansen and Collin \cite{HansenCollin_ANewChuFormulaForQ}  also included internal energies, but they used $\MAT{E}$- and $\MAT{H}$-fields together with the subtraction of the radial power-flow.  Hansen, Kim and Breinbjerg \cite{HansenKimBreinbjerg_StoredEnergyAndQualityFactorOnSphere} generalized the results of Hansen and Collin for any spherical TM and TE mode and for a sphere filled with an isotropic medium. The limitations of the dual mode case were studied by Fante \cite{Fante_QFactorOfGeneralIdeaAntennas} and recently by Kim \cite{Kim_MinimumQESA}.

The most recent approaches to the $\Q$ factor calculation utilized the source current distribution. There are obvious benefits: the resultant functionals are of bilinear forms, the calculation is very effective and it is possible to use any current distribution that is available thanks to modern EM simulators or that could even be user-defined. This opens new possibilities in optimization \cite{Gustafsson_OptimalAntennaCurrentsForQsuperdirectivityAndRP} and modal decomposition \cite{CapekHazdraEichler_AMethodForTheEvaluationOfRadiationQBasedOnModalApproach}. The work by Vandenbosch \cite{Vandenbosch_ReactiveEnergiesImpedanceAndQFactorOfRadiatingStructures} is inspired by the pioneering research of Geiy \cite{Geyi_AMethodForTheEvaluationOfSmallAntennaQ}, and directly uses Maxwell equations and the source currents. The same theory has been generalized in the time domain \cite{Vandenbosch_RadiatorsInTimeDom1}. However, some non-observable terms \cite{Rhodes_ObservableStoredEnergiesOfElectromagneticSystems} were neglected. Another approach by Gustafsson, Sohl and Kristensson \cite{GustafssonSohlKristensson_PhysicalLimitationsOfAntennasOfArbitraryShape_RoyalSoc} utilized static polarizability. Gustaffson and Jonsson \cite{Gustaffson_StoredElectromagneticEnergy_arXiv} also postulated the uncertainty in Vandenbosch's definition of $\Q$. Unfortunately, their contribution opens a new question about the coordinate dependent term which is strictly non-physical. 

Some attempts have also been made to obtain the lower bound of $\Q$ by utilizing the sources. This limit was investigated by Vandenbosch and Volski \cite{VandenboschVolski_LowerBoundsForRadiationQofVerySmallAntennasOfArbTopology}, but the method is encumbered with the difficulties mentioned above, and thus the results are provided only for a small radiator. Very interesting work has been done by Seshadri \cite{Seshadri_ResonancesOfASphericalAntenna}, closely related with \cite{Seshadri_ConstituentsOfPowerOfAnElectricDipoleOfFiniteSize}, where the complex power of the spherical modes is already known analytically.

Together with the theoretical achievements, many scientists have sought for an antenna prototype that achieves the given limits, see e.g. \cite{Sievenpiper_ExpretimentalValidationOfPerformanceLimitsAndDesignGuidelines}, \cite{BestHanna_AperformanceComparisonOfFundamentalESA}. The folded multi-arm spherical helix antenna designed by Best \cite{Best_TheRadiationPropertiesOfESAsphericalHelix} achieved roughly 1.5 times the Chu's limit and almost exactly the limit predicted by Hansen and Collin. An attempt to reach the Chu's limit was undertaken by Kim and Breinbjerg \cite{KimBreinbjerg_ReachingTheChuLowerBoundOnQwithMagneticDipoleAntennasUsingMagneticCoatedPECCore}, using a magnetic-coated PEC core.

The above mentioned history however evoke a question, whether the classical $\Q$ limits, based on the far-field energy extraction, are the only possibility how to establish an upper bound of FBW or whether there exist a simpler way. In fact, there exists another widely used concept of so-called $Q_Z$ factor proposed by Yaghjian and Best \cite{YaghjianBest_ImpedanceBandwidthAndQOfAntennas}, which should closely follow an inverse proportionality to FBW. Its source concept is already established \cite{CapekJelinekHazdraEichler_MeasurableQ}, however works on its lower bounds are scarce \cite{GustafssonNordebo_BandwidthQFactorAndResonanceModelsOfAntennas,Gustaffson_StoredElectromagneticEnergy_arXiv}. Particularly \cite{GustafssonNordebo_BandwidthQFactorAndResonanceModelsOfAntennas}, there exists an explicit evaluation of the $Q_Z$ of the separated TE and TM spherical modes with internal region excluded and there are signs of $Q_Z$ not having an absolute lower bound other than $Q_Z = 0$.

This paper makes amendments to the current state of the topic of the lower bounds of the $Q_Z$ factor. The method of $Q_Z$ evaluation is based on the differentiation of the complex power expressed by electromagnetic potentials rather than fields \cite{CapekJelinekHazdraEichler_MeasurableQ}. In this way, the issues with divergent integrals \cite{Vandenbosch_ReactiveEnergiesImpedanceAndQFactorOfRadiatingStructures} are automatically eliminated, since the subtraction of the far-field energy is not needed (it is implicitly included in the $Q_Z$ definition). The complex power differentiation also avoids non-physical quantities like coordinate dependent terms or negative energies \cite{Gustaffson_StoredElectromagneticEnergy_arXiv}.

The proposed theory is presented on an example of spherical modes, which have been in the spotlight in recent decades for their ability to establish a general lower bound of $\Q$ factor. It is important to stress that the whole process is completely analytical, without any approximations. The final expressions, presented in the closed form, are easy to work with and are compatible with all previous observations. Furthermore, the proposed methodology can be applied not only to the spherical coordinate system, but to any system in which the vector wave equation is separable \cite{MorseFeshBach_MethodsOfTheoreticalPhysics} and thus the vector potential is analytically known. This gives a possibility of practical  $Q_Z$  limits tailored for a particular antenna design.

The paper is organized as follows. The definition of the  $Q_Z$ factor  is briefly recapitulated in Section~\ref{Qdefinition}. The complex power and all necessary power and energy terms of the dominant spherical TM and TE modes are presented in Section~\ref{SphereTM} and Section~\ref{SphereTE}. Section~\ref{Limitations} presents the practically available limits  for single-mode radiators in free space, which are represented by the dominant spherical modes, further denoted as \TM and \TE and  compares them with the classical $\Q$ limits. Section~\ref{Sec_Combination} deals with the $Q_Z$ factor of the combination of the modes and compares it with the results for the classical $Q$ factor. Section~\ref{Sec_Discussion} then gives some important remarks on the presented derivations and results. The paper is concluded in Section~\ref{Sec_Concl}.

\section{Definition of  $Q_Z$ }
\label{Qdefinition}

The exact derivation of the  $Q_Z$  factor \cite{YaghjianBest_ImpedanceBandwidthAndQOfAntennas} in terms of sources is provided in \cite{CapekJelinekHazdraEichler_MeasurableQ}, including the related discussion and numerical verification, and reads
\BE
\label{EqX1}
\begin{split}
\QZc &= \left| \QRc + \J \QXc \right| \\
&= \frac{ka}{2 \left(\PM-\PE\right)} \left| \frac{\partial \Big(\left(\PM-\PE\right) + \J \omega \left(\WM-\WE\right)\Big)}{\partial ka } \right|,
\end{split}
\EE
 where subscripts $Z$, $R$ and $X$ represent impedance, resistance and reactance of the antenna respectively, and where $\J = \sqrt{-1}$, $\omega$ is the angular frequency of the time harmonic field \cite{Harrington_TimeHarmonicElmagField} under the convention \mbox{${\boldsymbol{\mathcal{F}}}\left( t \right) = \sqrt{2} \Re \left\{ \mathbf{F} \left( \omega \right) {\mathrm{e}}^{\J \omega t} \right\}$}, where $\boldsymbol{\mathcal{F}}$ is any time-harmonic quantity, $k = \omega / c_0$ is the wavenumber, $c_0$ is the speed of light, $a$ is the smallest radius of a sphere circumscribing all the sources, $\PM-\PE$ is the total radiated power \cite{Jackson_ClassicalElectrodynamics}, $\omega\left(\WM-\WE\right)$ is the total reactive power \cite{Jackson_ClassicalElectrodynamics}, and the total input current at the antenna's port is normalized to $I_0 = 1$A. Considering an arbitrary source current distribution $\MAT{J}$ and charge density $\rho$ inside a source region $\Omega$, and $\MAT{A}$ and $\varphi$ as the vector and scalar potential \cite{Jackson_ClassicalElectrodynamics}, the separated  $Q_R$ and $Q_X$ terms  in (\ref{EqX1}) can be written as 
\BE
\label{EqX2}
\QRc = \frac{\PM + \PE + \Prad + \Pom}{2 \left( \PM - \PE \right)},
\EE
and
\BE
\label{EqX3}
\QXc = \frac{\omega \left( \WM + \WE + \Wrad + \Wom \right)}{2 \left( \PM - \PE \right)},
\EE
where the particular terms are expressed as
\begin{subequations}
\begin{align}
\label{EqX4_A}
\WM - \J \frac{\PM}{\omega} &= \int\limits_\Omega \MAT{A} \cdot \MAT{J}^\ast \D{\mathbf{r}}, \\
\label{EqX4_B}
\WE - \J \frac{\PE}{\omega} &= \int\limits_\Omega \varphi \rho^\ast \D{\mathbf{r}}, \\
\label{EqX4_C}
\Wrad - \J \frac{\Prad}{\omega} &=  - \J k \left( k^2 \mathcal{L}_{\mathrm{rad}} \left( \mathbf{J}, \mathbf{J} \right) - \mathcal{L}_{\mathrm{rad}} \left( \nabla \cdot \mathbf{J}, \nabla \cdot \mathbf{J} \right) \right), \\
\label{EqX4_D}
\Wom - \J \frac{\Pom}{\omega} &=  k^2 \mathcal{L}_{\omega} \left( \mathbf{J}, \mathbf{J} \right) - \mathcal{L}_{\omega} \left( \nabla \cdot \mathbf{J}, \nabla \cdot \mathbf{J} \right),
\end{align}
\end{subequations}
with
\begin{subequations}
\begin{align}
\label{EqX5_B}
\mathcal{L}_{\mathrm{rad}} \left( \mathbf{U}, \mathbf{V} \right) &= \frac{1}{4 \pi \epsilon \omega^2} \int\limits_{\Omega '} \int\limits_{\Omega} \mathbf{U} \left( \mathbf{r} \right) \cdot \mathbf{V}^\ast \left( \mathbf{r} ' \right) \mathrm{e}^{-\J k R} \, \mathrm{d} \mathbf{r} \, \mathrm{d} \mathbf{r} ' , \\
\label{EqX5_C}
\mathcal{L}_{\omega} \left( \mathbf{U}, \mathbf{V} \right) &= \frac{1}{4 \pi \epsilon \omega} \int\limits_{\Omega '} \int\limits_{\Omega}  \frac{\partial \left( \mathbf{U} \left( \mathbf{r} \right) \cdot \mathbf{V}^\ast \left( \mathbf{r} ' \right) \right)}{\partial\omega} \frac{\mathrm{e}^{-\J k R}}{R} \, \mathrm{d} \mathbf{r} \, \mathrm{d} \mathbf{r} ',
\end{align}
\end{subequations}
in which \mbox{$R = \left\| \mathbf{r} - \mathbf{r} ' \right\|$} is the Euclidean distance, $\epsilon$ is the vacuum permittivity, and $\ast$ denotes complex conjugation. The detailed derivation of the above relations is described in \cite{CapekJelinekHazdraEichler_MeasurableQ}.

The $\QXc$ in (\ref{EqX3}) and $\QZc$ in (\ref{EqX1}) are $\Q$ factors of an untuned antenna \cite{VolakisChenFujimoto_SmallAntennas_MiniatrurizTechniques} and thus they will be denoted as $\QXcU$ and $\QZcU$ in the rest of the paper. One can, however, tune the antenna to its resonance at angular frequency $\omega_0$ by a reactive lumped element. Then,  the $\Q$ factors of tuned antenna  will be denoted as $\QXcT$ and $\QZcT$, and they can be evaluated as \cite{CapekJelinekHazdraEichler_MeasurableQ}
\begin{subequations}
\begin{align}
\label{EqX6}
\QXcT &= \omega_0 \frac{2 \max\left\{ W_{\mathrm{m}}, W_{\mathrm{e}}\right\} + W_{\mathrm{rad}} + W_\omega}{2 \left( P_{\mathrm{m}} - P_{\mathrm{e}} \right)}, \\
\label{EqXB}
\QZcT &= \left|Q_R + \J \QXcT\right|.
\end{align}
\end{subequations}
Note that tuning by purely reactive elements leaves the $\QRc$ factor unchanged.

\section{Complex power and  the $Q_Z$  of the \TM mode}
\label{SphereTM}

Let us consider the \TM mode, which is described by the current density
\BE
\label{EqTM1}
\MAT{J} = \frac{\sin\left(\vartheta \right)}{2 \pi a} \delta\left( r - a \right) \boldsymbol\vartheta_0
\EE
flowing on a spherical shell of radius $a$ situated in a vacuum, where $\delta$ is the Dirac delta and $\boldsymbol\vartheta_0$ is the unit vector co-directional with $\vartheta$, see Fig.~\ref{fig:modes}.
\BF
\includegraphics[width=9.2cm]{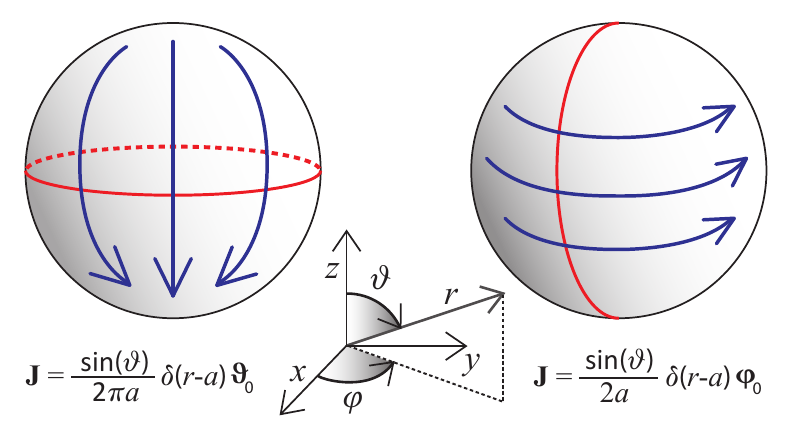}
\caption{Sketch of modal currents on a sphere of radius $a$: dipole mode (\TM) on the left and loop mode (\TE) on the right. The input current is normalized with respect to the red contours. The coordinate system considered throughout the paper is depicted in the middle of the figure.}
\label{fig:modes}
\EF
The current density (\ref{EqTM1}) is normalized so that the current flowing through the  $x$-$y$  plane is $I_0 = 1$A. The corresponding charge density is
\BE
\label{EqTM2}
\rho = \J \frac{\cos\left(\vartheta \right)}{\omega \pi a^2} \delta\left( r - a \right).
\EE

The vector and scalar potentials of the \TM mode are (see appendix~\ref{app2})
\BE
\label{EqTM5}
\begin{split}
A_\vartheta = &-\frac{\J \mu}{2 \pi ka} \sin \left( \vartheta \right) \Big( 2 \MR{h}_1^{(2)} \left( ka\right) \MR{j}_1 \left( ka\right) \\
&+ \left( \MR{h}_1^{(2)} \left( ka\right) - ka\, \MR{h}_0^{(2)} \left( ka\right) \right) \left( \MR{j}_1 \left( ka\right) - ka \, \MR{j}_0 \left( ka\right) \right) \Big) 
\end{split}
\EE
and
\BE
\label{EqTM6}
\varphi = \frac{\omega\mu}{\pi k} \MR{h}_1^{(2)} \left( ka\right) \MR{j}_1 \left( ka\right) \cos \left( \vartheta \right),
\EE
where $\MR{j}_n$ and $\MR{h}_n^{(2)} = \MR{j}_n - \J \MR{y}_n$ are the spherical Bessel and Hankel functions of the $n$th order \cite{Jeffrey_MathHandbook}. Substituting the potentials into (\ref{EqX4_A}) and (\ref{EqX4_B}) leads to
\begin{subequations}
\begin{align}
\label{EqTM7_A}
\PM =& \frac{4}{6 \pi} Z_0 \left( 2 \MR{j}_1^2 \left( ka\right) + \big( \MR{j}_1 \left( ka\right) - ka \, \MR{j}_0 \left( ka\right) \big)^2 \right), \\
\label{EqTM7_B}
\PE =& \frac{4}{3 \pi} Z_0 \MR{j}_1^2 \left( ka\right), \\
\label{EqTM7_C}
\omega\WM = &-\frac{4}{6 \pi} Z_0 \Big( 2 \MR{y}_1 \left( ka\right) \MR{j}_1 \left( ka\right) \\
&+ \big( \MR{y}_1 \left( ka\right) - ka \, \MR{y}_0 \left( ka\right) \big) \big( \MR{j}_1 \left( ka\right) - ka \, \MR{j}_0 \left( ka\right) \big) \Big), \nonumber \\
\label{EqTM7_D}
\omega\WE = & -\frac{4}{3 \pi} Z_0 \MR{y}_1 \left( ka\right) \MR{j}_1 \left( ka\right),
\end{align}
\end{subequations}
where $Z_0 = \sqrt{\mu / \epsilon}$ is the free space impedance. Note here that the distribution (\ref{EqTM1}), by definition, does not vary with the frequency, $\partial \MAT{J} \left( \vartheta \right) / \partial \omega = 0$, and thus from (\ref{EqX4_D}) we have
\BE
\label{EqTM3}
\Pom = \omega \Wom = 0.
\EE
Finally, by comparing (\ref{EqX1}) with (\ref{EqX2}) and (\ref{EqX3}), and using (\ref{EqTM3}), we can deduce that
\begin{subequations}
\begin{align}
\label{EqTM4_A}
\Prad &= ka \frac{\partial \left( \PM - \PE \right)}{\partial ka} - \left( \PM + \PE \right), \\
\label{EqTM4_B}
\omega \Wrad &= ka \frac{\partial \omega \left( \WM - \WE \right)}{\partial ka} - \omega \left( \WM + \WE \right).
\end{align}
\end{subequations}
The above expressions have been simplified in Mathematica \cite{Mathematica} and evaluated in Matlab \cite{Matlab}, and the results are depicted in Fig.~\ref{fig:Pterms_TM}. The corresponding $\QRc$, $\QXcU$ and $\QZcU$ factors of the \TM mode are depicted in Fig.~\ref{fig:Quntuned_TM}.

\BF
\includegraphics[width=9.2cm]{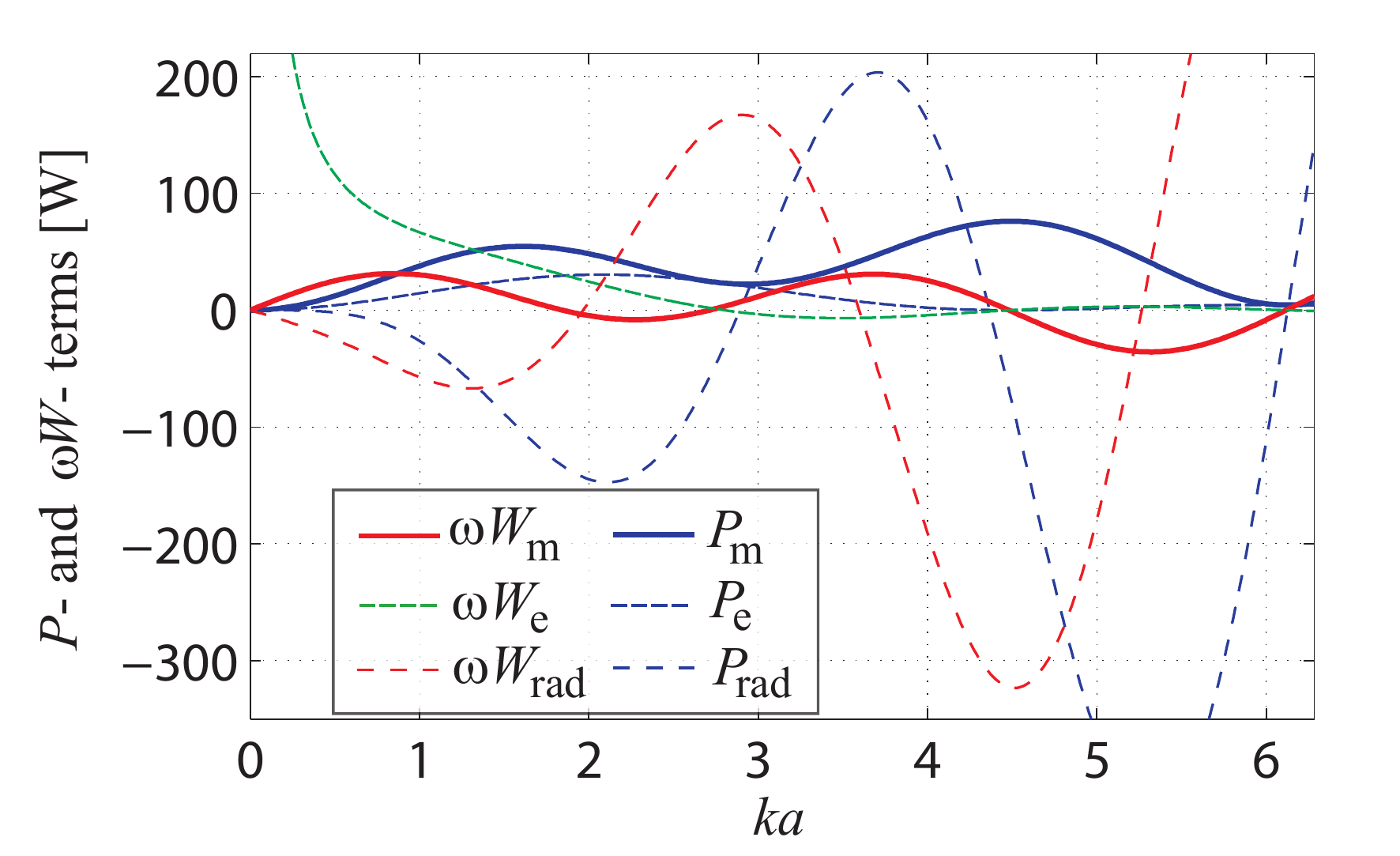}
\caption{The radiated power terms $\PM$ and $\PE$, the reactive power terms $\omega\WM$ and $\omega\WE$, and the power terms associated with radiation $\Prad$ and $\omega\Wrad$ for the \TM mode.}
\label{fig:Pterms_TM}
\EF

\BF
\includegraphics[width=9.2cm]{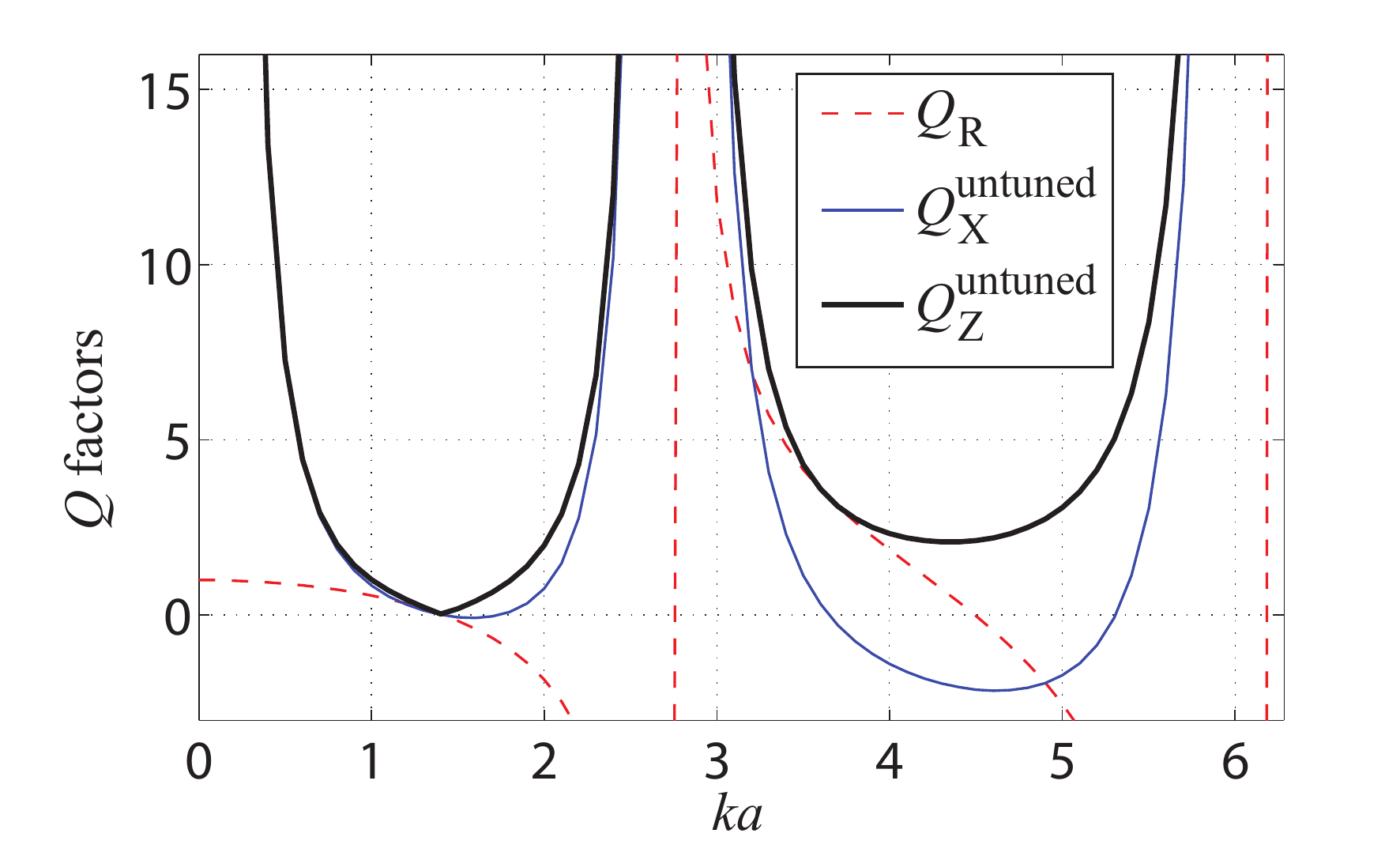}
\caption{ Comparison of the $\QZcU$ and its parts $\QRc$ and $\QXcU$ for the \TM mode.  Both $\QRc$ and $\QXcU$ can be negative. They can only be interpreted according to (\ref{EqX1}).}
\label{fig:Quntuned_TM}
\EF

\section{Complex power and the $Q_Z$ of the \TE mode}
\label{SphereTE}

The procedure from the previous section can be used for the \TE mode as well. In that case, the current density is
\BE
\label{EqTE1}
\MAT{J} = \frac{\sin\left(\vartheta \right)}{2 a} \delta\left( r - a \right) \boldsymbol\varphi_0,
\EE
where $\boldsymbol\varphi_0$ is the unit vector co-directional with $\varphi$, see Fig.~\ref{fig:modes}. The current density is normalized so that the current flowing through the  $z$-$\left( x>0 \right)$  half-plane is $I_0 = 1$A. The corresponding charge density vanishes, $\rho = 0$, and so
\begin{subequations}
\begin{align}
\label{EqTE2_A}
\varphi =& 0, \\
\label{EqTE2_B}
\PE =& 0, \\
\label{EqTE2_C}
\omega\WE = & 0.
\end{align}
\end{subequations}
Furthermore, as the current is frequency independent, (\ref{EqTM3})--(\ref{EqTM4_B}) still holds. The vector potential is again found by the method described in appendix~\ref{app2}, and is equal to
\BE
\label{EqTE3}
A_\varphi = -\frac{\J \mu}{2} \sin\left(\vartheta\right) ka \, \MR{j}_1 \left( ka \right) \MR{h}_1^{(2)} \left( ka \right),
\EE
which leads to
\begin{subequations}
\begin{align}
\label{EqTE4_A}
\PM =& \frac{2 \pi}{3} Z_0 \left( ka \right)^2 \MR{j}_1^2 \left( ka \right), \\
\label{EqTE4_B}
\omega\WM = & -\frac{2 \pi}{3} Z_0 \left( ka \right)^2 \MR{j}_1 \left( ka \right) \MR{y}_1 \left( ka \right).
\end{align}
\end{subequations}

All non-zero terms related to the \TE mode are depicted in Fig.~\ref{fig:Pterms_TE}. The $\QZcU$ and its parts $\QRc$ and $\QXcU$  of the \TE mode  are depicted in Fig.~\ref{fig:Quntuned_TE}.

\BF
\includegraphics[width=9.2cm]{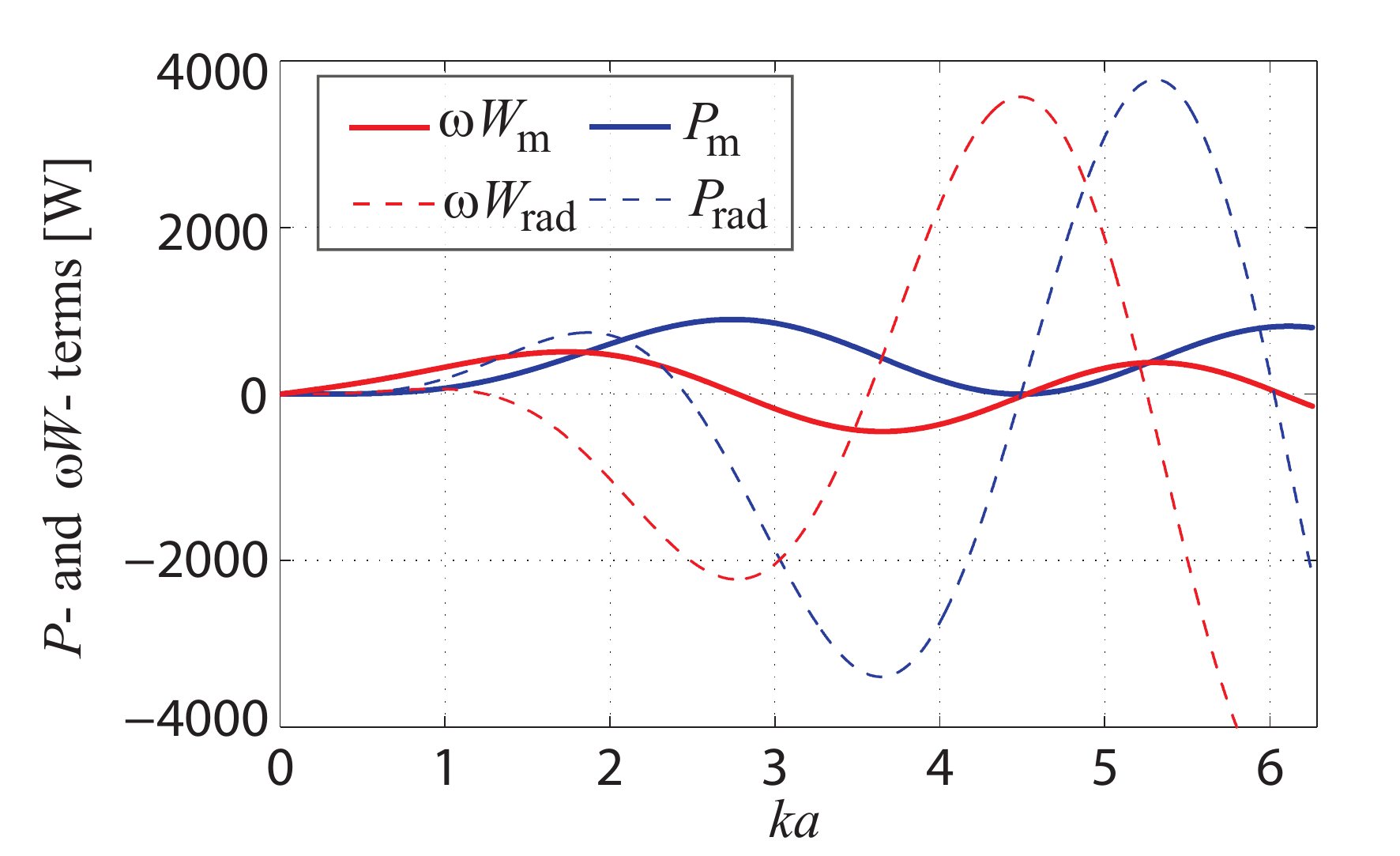}
\caption{The radiated power term $\PM$, the reactive power term $\omega\WM$, and the power terms associated with radiation $\Prad$ and $\omega\Wrad$ for the \TE mode.}
\label{fig:Pterms_TE}
\EF

\BF
\includegraphics[width=9.2cm]{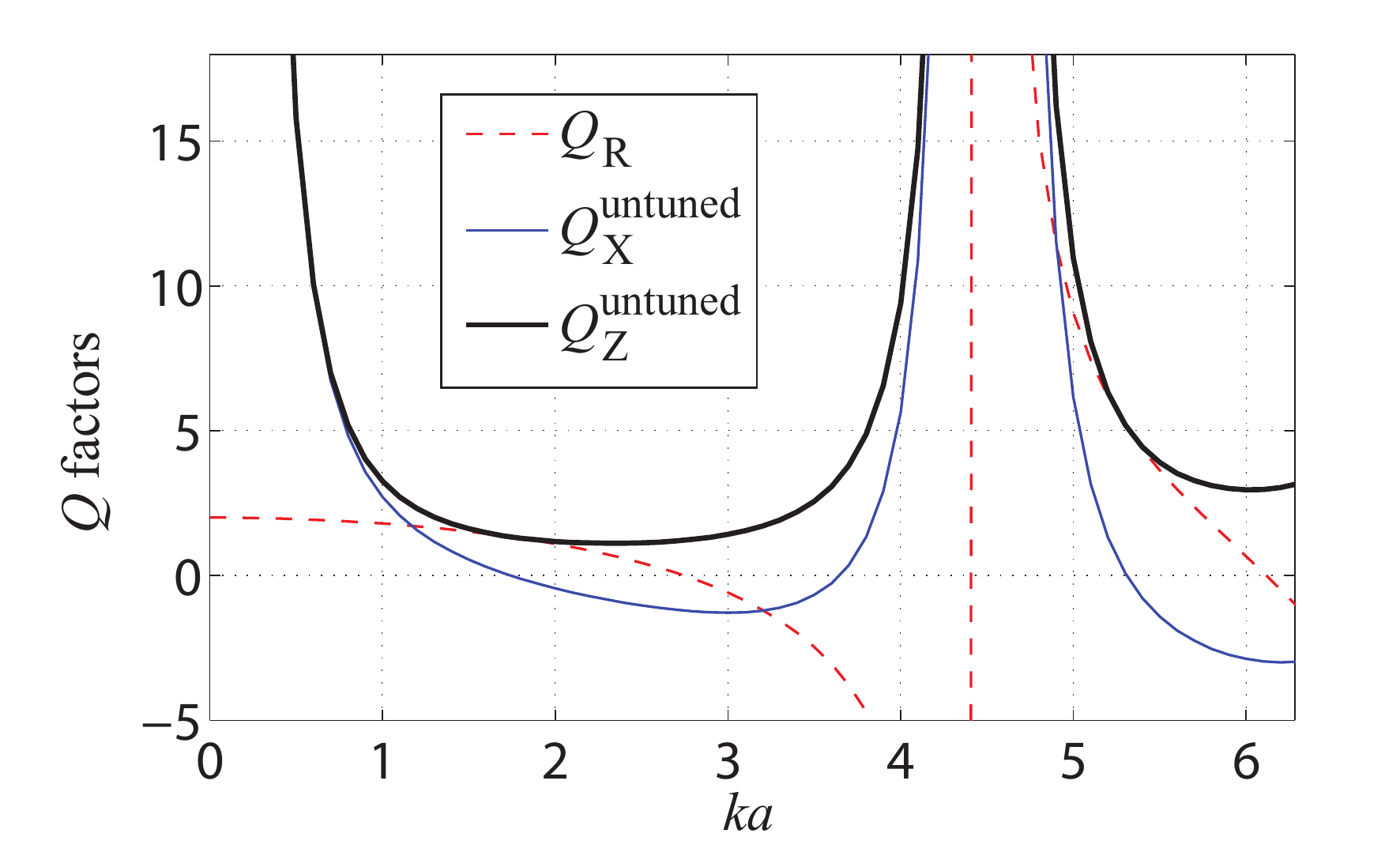}
\caption{Comparison of the $\QZcU$ and its parts $\QRc$ and $\QXcU$ for the \TE mode.  Both $\QRc$ and $\QXcU$ can be negative. They can only be interpreted according (\ref{EqX1}).}
\label{fig:Quntuned_TE}
\EF

\section{The limitations for ESA and asymptotic behaviour of  the $Q_Z$ factor  for the \TM and the \TE mode}
\label{Limitations}

In this section, we will discuss the $\Q$  and the $Q_Z$  factors for the spherical \TM and \TE modes that are tuned to its resonance at given $ka$ by the external reactive lumped element. Particularly, the $\QZcT$ obtained from (\ref{EqXB}) is compared with the classical Chu's limit \cite{Chu_PhysicalLimitationsOfOmniDirectAntennas} $\Qcollin$ (formula (8) in \cite{CollinRotchild_EvaluationOfAntennaQ}), with the limit found by Hansen and Collin \cite{HansenCollin_ANewChuFormulaForQ} $\Qhansen$  (formulas (9) and (12) of \cite{HansenCollin_ANewChuFormulaForQ}) and with the recent limits found by Vandenbosh \cite{Vandenbosch_ReactiveEnergiesImpedanceAndQFactorOfRadiatingStructures}. Note that within the context of frequency independent modes (with \mbox{$\Pom = \omega \Wom = 0$}) the quality factor used by Vandenbosh \cite{Vandenbosch_ReactiveEnergiesImpedanceAndQFactorOfRadiatingStructures} is just the $\QXcT$.

The results for the \TM mode are depicted in Fig.~\ref{fig:Qtuned_TM}, while the results for the \TE mode are depicted in Fig.~\ref{fig:Qtuned_TE}. The region of ESA ($ka < 0.5$), which is of interest for $\Q$ factor limits, is highlighted. The observed agreement between $\QZcT$, $\QXcT$ and $\Qhansen$ can be denoted as excellent in this region. On the other hand, the Chu's limit is clearly too optimistic due to the fact that Chu excluded the reactive power inside the bounding sphere. This is particularly significant in the case of the \TE mode, which stores  approximately one third of the total stored energy inside the sphere.

\BF
\includegraphics[width=9.2cm]{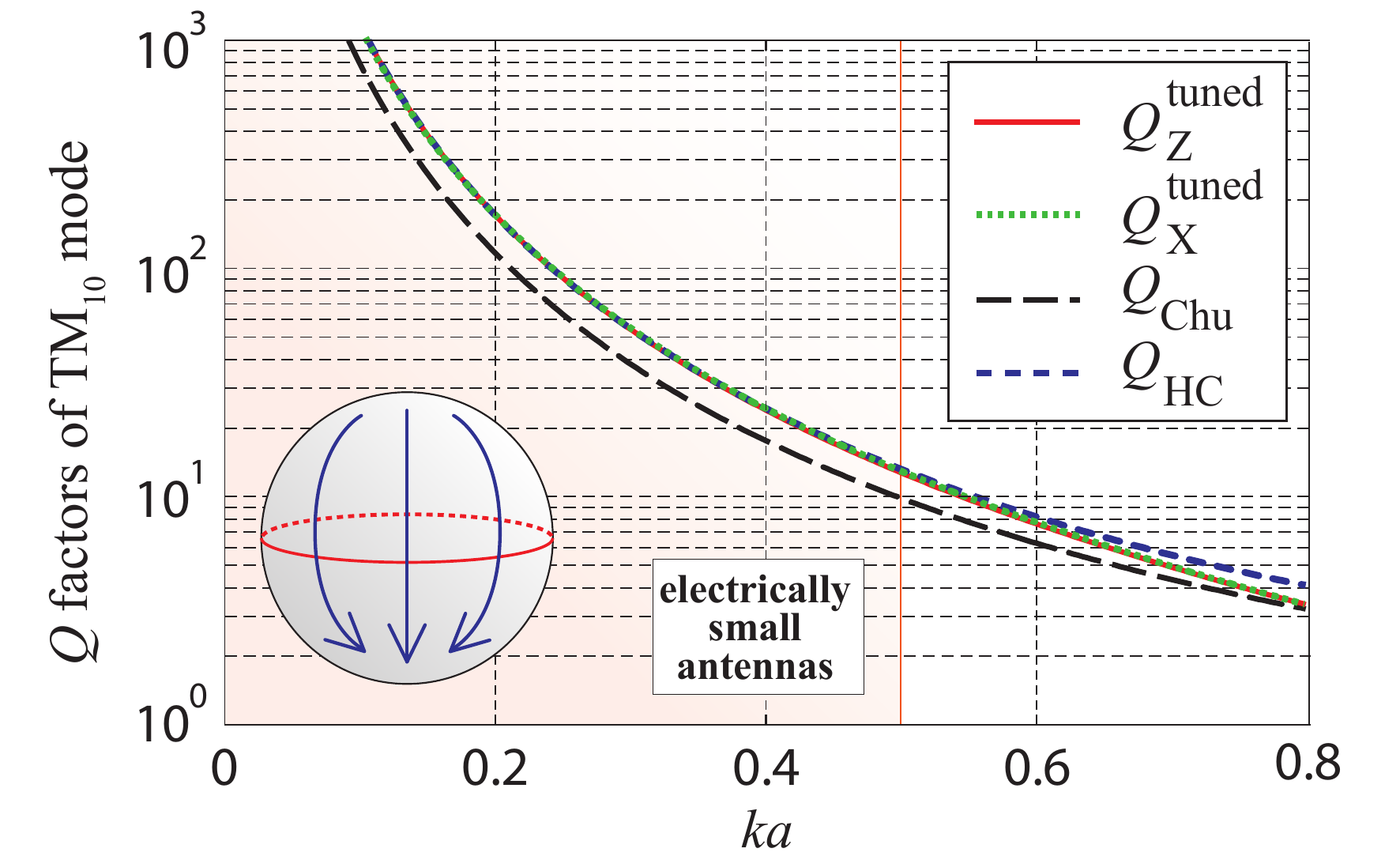}
\caption{Comparison of the $\Q$ and the $Q_Z$ factors for the spherical \TM mode: the $\QZcT$ of this paper, the $\Qcollin$ from \cite{CollinRotchild_EvaluationOfAntennaQ}, the $\Qhansen$ from \cite{HansenCollin_ANewChuFormulaForQ} and $\QXcT$ of this paper which is equivalent to that of \cite{Vandenbosch_ReactiveEnergiesImpedanceAndQFactorOfRadiatingStructures}.}
\label{fig:Qtuned_TM}
\EF
\BF
\includegraphics[width=9.2cm]{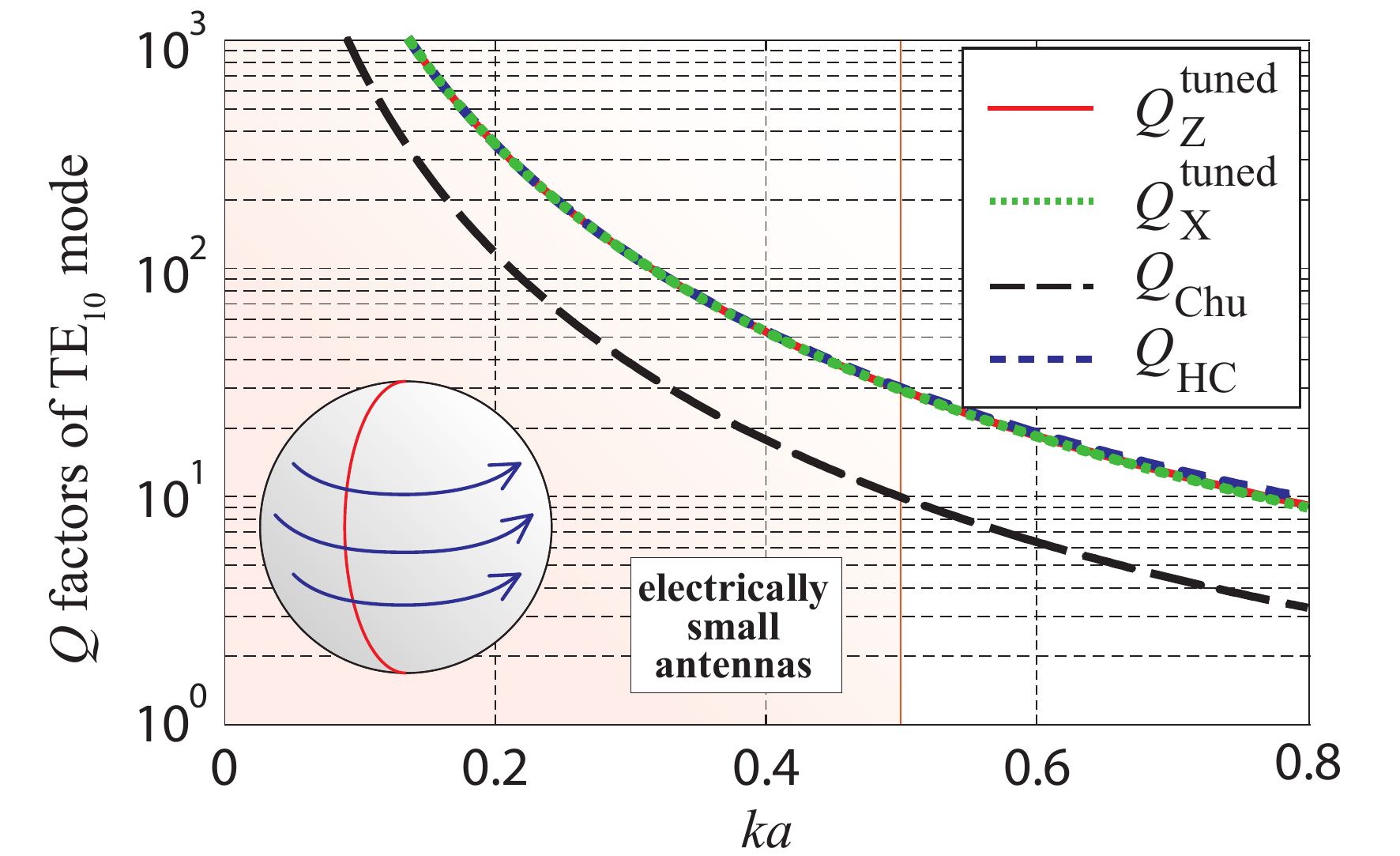}
\caption{Comparison of the $\Q$ and the $Q_Z$ factors for the spherical \TE mode: the $\QZcT$ of this paper, the $\Qcollin$ from \cite{CollinRotchild_EvaluationOfAntennaQ}, the $\Qhansen$ from \cite{HansenCollin_ANewChuFormulaForQ} and $\QXcT$ of this paper which is equivalent to that of \cite{Vandenbosch_ReactiveEnergiesImpedanceAndQFactorOfRadiatingStructures}.}
\label{fig:Qtuned_TE}
\EF

The explicit forms of the $\QZcT$ and $\Qhansen$ of the \TM and of the \TE mode are rather lengthy and are thus left for the Appendix~\ref{app3}. Within the region of interest (the ESA region of $ka < 0.5$) their series expansion provide an excellent approximation and the series expansion can furthermore be directly compared to the published results of the classical $\Q$ limits. As for the $Q_R$, $\QXcT$ and $\QZcT$ factors derived in this paper, the series for the \TM mode read
\begin{subequations}
\begin{align}
\label{EqLimits3A}
\QRc &= 1 - \frac{2}{5} \left(ka \right)^2 + \mathcal{O}\left( ka \right)^4,\\
\label{EqLimits3B}
\QXcT &= \frac{3}{2 \left( ka \right)^3} + \frac{3}{5 \left( ka \right)} - \frac{813}{1400}ka + \mathcal{O}\left( ka \right)^3,\\
\label{EqLimits3C}
\QZcT &=  \QXcT + \mathcal{O}\left( ka \right)^3,
\end{align}
\end{subequations}
while the series for the \TE mode read
\begin{subequations}
\begin{align}
\label{EqLimits4A}
\QRc &= 2 - \frac{\left( ka \right)^2}{5} + \mathcal{O}\left( ka \right)^4,\\
\label{EqLimits4B}
\QXcT &= \frac{3}{\left( ka \right)^3} + \frac{3}{ka} - \frac{174}{175}ka + \mathcal{O}\left( ka \right)^3,\\
\label{EqLimits4C}
\QZcT & =  \QXcT + \mathcal{O}\left( ka \right)^3.
\end{align}
\end{subequations}
The $\QZcT$ is almost identical to $\QXcT$ for \mbox{$ka \rightarrow 0$}, since \mbox{$\QXcT \gg \QRc$}, see Fig.~\ref{fig:Quntuned_TM} and Fig.~\ref{fig:Quntuned_TE}. 
For comparison we also present series of the classical Chu's limit \cite{Chu_PhysicalLimitationsOfOmniDirectAntennas} in the version of \cite{CollinRotchild_EvaluationOfAntennaQ} which reads
\BE
\label{EqLimits20}
\Qcollin = \frac{1}{\left( ka \right)^3} + \frac{1}{ka},
\EE
(note that this expansion is an exact formula) and the series of $\Qhansen$ of the \TM mode \cite{HansenCollin_ANewChuFormulaForQ} which reads
\BE
\label{EqLimits2A}
\Qhansen \approx \frac{3}{2 \left(ka \right)^3} + \frac{1}{\sqrt{2} ka}.
\EE
Lastly, the $\Qhansen$ of the \TE mode \cite{HansenCollin_ANewChuFormulaForQ} reads
\BE
\label{EqLimits2B}
\Qhansen \approx 3 \Qcollin.
\EE
Comparing the above expressions, a good correspondence between (\ref{EqLimits2A}), (\ref{EqLimits3C}) and (\ref{EqLimits2B}), (\ref{EqLimits4C}) is now evident. 

\section{The $Q_Z$ factor of the Linear Combination of the \TM and the \TE mode}
\label{Sec_Combination}

Let us now assume more complicated example consisting of a linear combination of collinear magnetic and electric dipole forming a generalized Huygens source. In such case, the surface current density on the spherical shell can be expressed as
\BE
\label{NewEq1}
{\mathbf{J}} = {{\bf{J}}_{{\rm{TM}}}} + Y\left( ka  \right){{\bf{J}}_{{\rm{TE}}}},
\EE
where ${{\bf{J}}_{{\rm{TM}}}}$ and ${{\bf{J}}_{{\rm{TE}}}}$ are given by (\ref{EqTM1}) and (\ref{EqTE1}), respectively, i.e. the current normalization is performed over the \TM mode only. As formulated, the coefficient $Y\left( ka  \right)$ has to represent a transfer function of a causal system (being analytical in the lower half-plane of complex $k$), but otherwise can be arbitrary.

The current density (\ref{NewEq1}) is frequency dependent and thus (\ref{EqTM3}), (\ref{EqTM4_A}), (\ref{EqTM4_B}) are not valid any more. We can however follow a similar scheme and by comparing (\ref{EqX1}) with (\ref{EqX2}) and (\ref{EqX3}), and using (\ref{EqX6}) and (\ref{EqXB}) we can easily deduce that
\BE
\label{NewEq2}
Q_Z^{{\rm{tuned}}} = \,\frac{1}{{2\left( {{P_{\rm{m}}} - {P_{\rm{e}}}} \right)}}\left| {ka \frac{{\partial \Big( {\left( {{P_{\rm{m}}} - {P_{\rm{e}}}} \right) + \jmath \omega \left( {{W_{\rm{m}}} - {W_{\rm{e}}}} \right)} \Big)}}{{\partial ka }} + \jmath \omega \left| {{W_{\rm{m}}} - {W_{\rm{e}}}} \right|} \right|,
\EE
which could be taken as a general formula for the $\QZcT$ evaluation. The resulting $\QZcT$ will also be compared with the $\Qhansen$ evaluated by the classical extraction method of Collin and Rothschild \cite{CollinRotchild_EvaluationOfAntennaQ} (refined by Hansen and Collin \cite{HansenCollin_ANewChuFormulaForQ}). 

The final formulas are rather clumsy in both cases to be shown explicitly and we will thus stick only to graphical results. In this respect, it is very interesting to use the new degree of freedom gained by the coefficient $Y\left( ka  \right)$ in (\ref{NewEq1}). More specifically, it is straightforward to show that the $\Qhansen$ depends solely on ${\left| {Y\left( ka  \right)} \right|^2}$, while the $\QZcT$ depends on both, ${\left| {Y\left( ka  \right)} \right|^2}$ and \mbox{$\partial \left| {Y\left( ka  \right)} \right|^2 /\, \partial ka$}. The $\QZcT$ thus, in fact, gained two new degrees of freedom. With respect to this paper it is then interesting to ask whether the coefficient $Y\left( ka  \right)$ and its derivative cannot be optimized so that the $\QZcT$ and the $\Qhansen$ of the combined current (\ref{NewEq1}) would, at a given $ka$, reach lower values than the $\QZcT$ and the $\Qhansen$ of the pure \TM mode. The results are depicted in Fig.~\ref{fig:Qoptim_QHC} and Fig.~\ref{fig:Qoptim_QZ} for minimized values of the $Q$ factors at four different values of $ka$. As a word of caution it is important to note here that the curves in Fig.~\ref{fig:Qoptim_QHC} and Fig.~\ref{fig:Qoptim_QZ} were obtained in such a way that the $\QZcT$ and the $\Qhansen$ were optimized at a selected value of $ka$ by the variation of ${\left| {Y\left( ka  \right)} \right|^2}$ and \mbox{$\partial \left| {Y\left( ka  \right)} \right|^2 /\, \partial ka$} and the optimal values of the variables were then kept in the rest of the depicted $ka$ interval. This represents no problem for the $\Qhansen$, which depends solely on ${\left| {Y\left( ka  \right)} \right|^2}$, but can only be approximately satisfied in the case of $\QZcT$, which depends on both, ${\left| {Y\left( ka  \right)} \right|^2}$ and \mbox{$\partial \left| {Y\left( ka  \right)} \right|^2 /\, \partial ka$}.

The curves in Fig.~\ref{fig:Qoptim_QHC} and Fig.~\ref{fig:Qoptim_QZ} clearly show that both, the $\Qhansen$ and the $\QZcT$ can be lowered by a proper combination of the \TM and the \TE modes. The results for the $\Qhansen$ are coherent with \cite{2009_Thal_TAP,Pozar_NewResultsForMinimumQMaximumGainAndPolarization} and show just a mild drop. On the other hand, the $\QZcT$ can be cast to values as low as \mbox{$\QZcT \approx 1$}, which, against all odds, can be done even for $ka \rightarrow 0$. Moreover, it can be shown that for approximately \mbox{$ka > 1.3$}, the $\QZcT$ can always be made equal to zero, which follows the predictions of \cite{GustafssonNordebo_BandwidthQFactorAndResonanceModelsOfAntennas,Gustaffson_StoredElectromagneticEnergy_arXiv}.

\BF
\includegraphics[width=9.2cm]{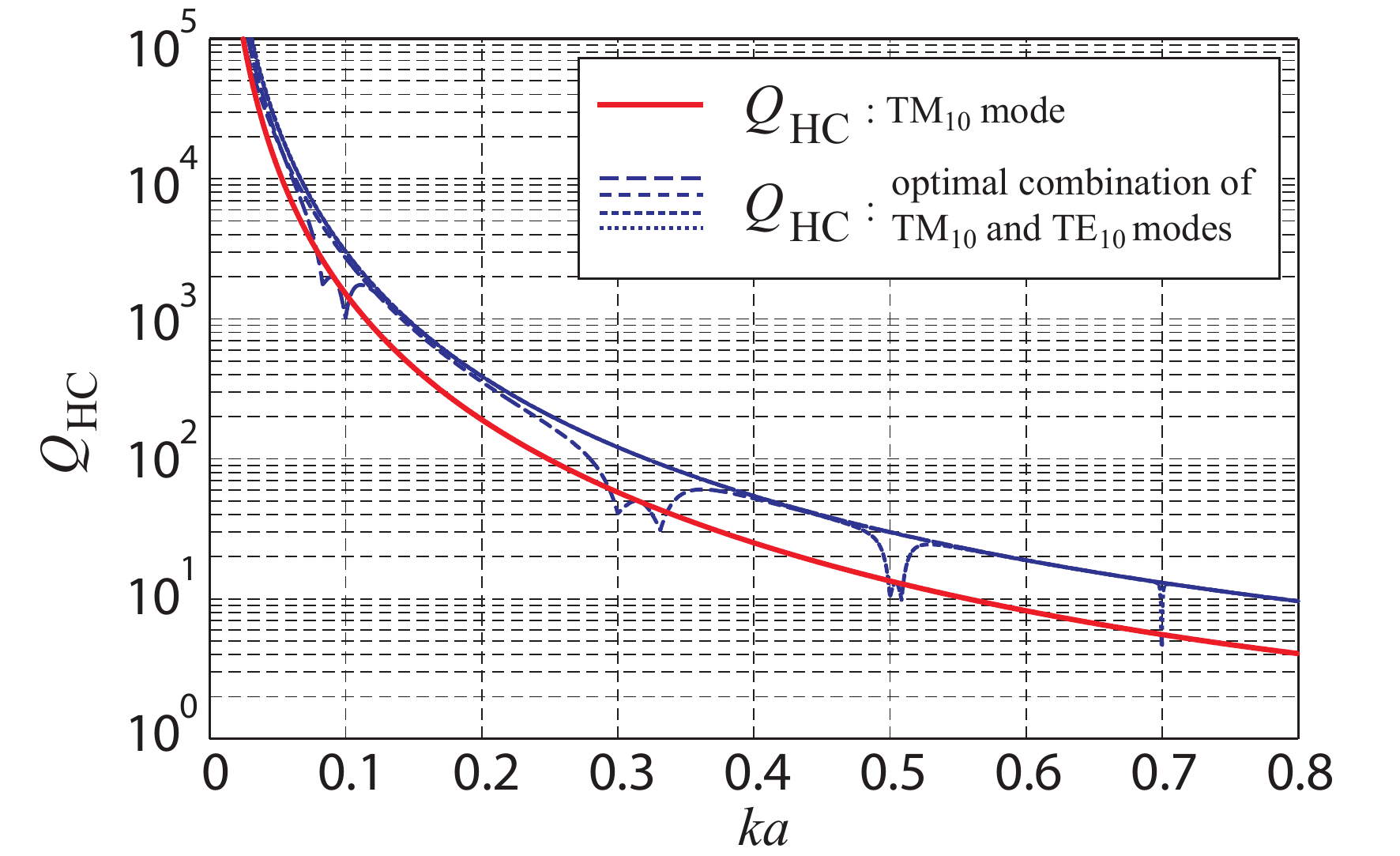}
\caption{Comparison of the $\Qhansen$ factor of a combination of the \TM and the \TE modes for several optimal realizations of $Y \left(ka \right)$ (blue dashed curves) and of the $\Qhansen$ of the pure \TM mode (red curve).}
\label{fig:Qoptim_QHC}
\EF

\BF
\includegraphics[width=9.2cm]{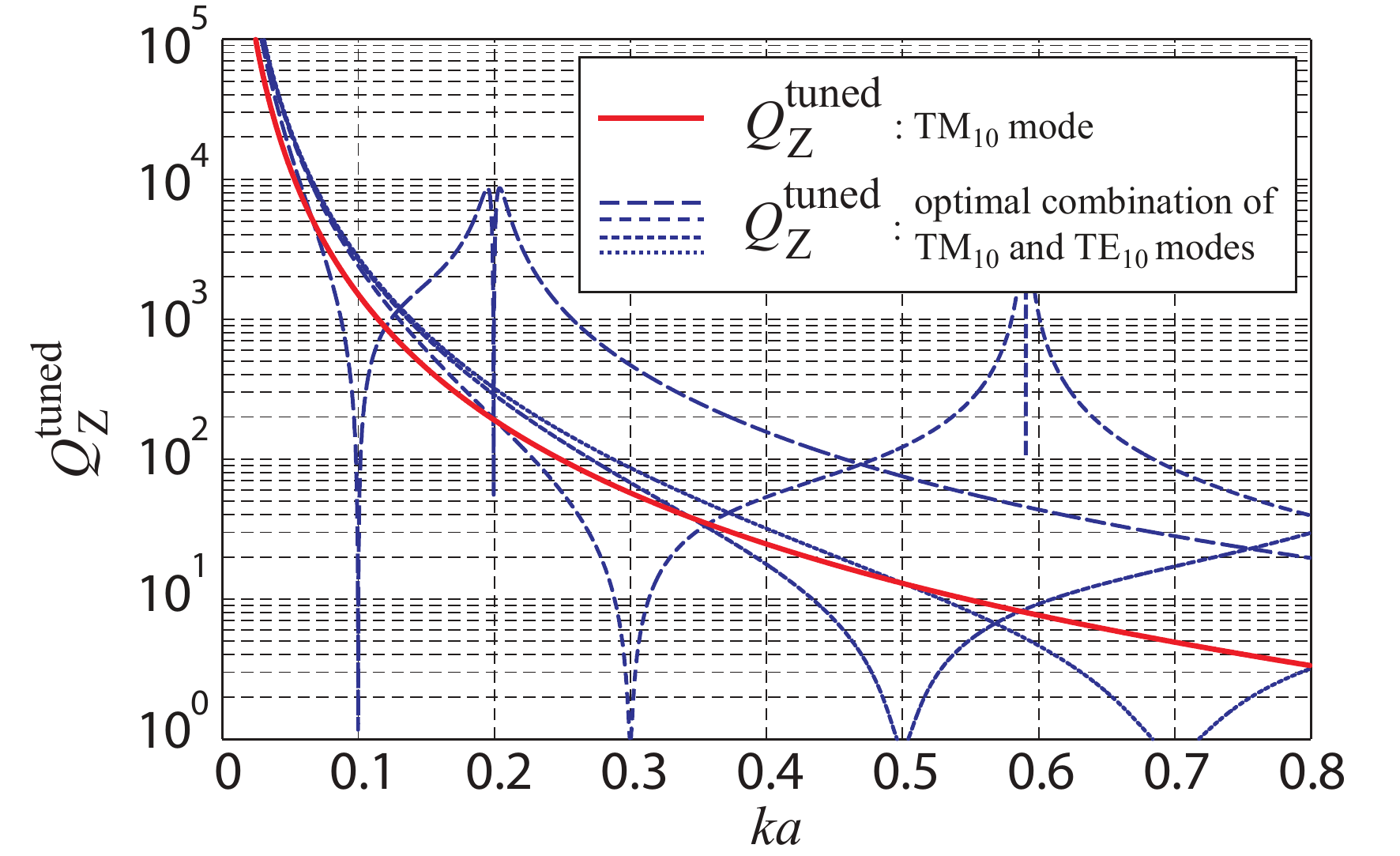}
\caption{Comparison of the $\QZcT$ factor of a combination of the \TM and the \TE modes for several optimal realizations of $Y \left(ka \right)$ (blue dashed curves) and of the $\QZcT$ of the pure \TM mode (red curve).}
\label{fig:Qoptim_QZ}
\EF


\section{Interpretation of the results}
\label{Sec_Discussion}

In this section, we will briefly comment and offer an interpretation for the above derived results in order to make the message of the paper perspicuous.
\begin{itemize}
	\item{The presented method describes a general scheme for the $\QZcT$ evaluation in separable coordinate systems \cite{Stratton_ElectromagneticTheory}. Since no explicit far-field energy extraction is needed, the method provides much simpler evaluation scheme than the classical derivation. In particular, the presented method avoids the complicated integration in the ``radial" direction. The only underlying integrals (\ref{EqX4_A}) and (\ref{EqX4_B}) can easily be evaluated in-hand for all common separable systems (spherical, cylindrical, spheroidal, ellipsoidal) since the integrands are polynomials of trigonometric functions. As a consequence, the evaluation of the $\QZcT$ can in principle be done analytically in the separable systems.}
			\item{The good correspondence of the $\QZcT$ with the classical $Q$ for the \TM and the \TE modes is reported for \mbox{$ka < 1$}. This fact has, up to now, been only known for the $\QZcT$ and the $Q$ referring to the fields external to the spherical shell of sources \cite{GustafssonNordebo_BandwidthQFactorAndResonanceModelsOfAntennas}, but has only been in the realm of a hypothesis for the fields including the internal region. The above given derivation puts this claim on solid grounds.}
			\item{The Section~\ref{Sec_Combination} reveals, following the predictions of \cite{GustafssonNordebo_BandwidthQFactorAndResonanceModelsOfAntennas,Gustaffson_StoredElectromagneticEnergy_arXiv}, that when dealing with the combination of at least two modes, the $\QZcT$ can become problematic. Although the $\QZcT$ of the combination of the \TM and the \TE mode is mostly higher than the $\QZcT$ of the pure \TM mode, it can be locally lowered. The reduction can be much stronger than $\QZcT / 2$ of the pure \TM mode, which would be expected for the classical $Q$ factor \cite{2009_Thal_TAP,Pozar_NewResultsForMinimumQMaximumGainAndPolarization}. The reduction can cover many orders in magnitude and the $\QZcT$ can in fact reach even zero value. Nevertheless, it should be recognized that this phenomenon does not reflect a physical reality of enhancing the FBW, but rather represents a shortcoming of the $\QZcT$. In practical designs of electrically small antennas, the $\QZcT$ of the pure \TM mode should still be considered as a reasonable lower bound.}
		\item{The previous observation implies that the following statement on the $Q_Z$ factor from the IEEE Std. \cite{IEEEStd_antennas}: ``\textit{NOTE - For an electrically small antenna, it is numerically equal to one-half the magnitude of the ratio of the incremental change in impedance to the corresponding incremental change in frequency at resonance, divided by the ratio of the antenna resistance to the resonant frequency.}" should be revisited, since the counter-example is provided in this paper and in \cite{GustafssonNordebo_BandwidthQFactorAndResonanceModelsOfAntennas,Gustaffson_StoredElectromagneticEnergy_arXiv}.}

	\item{Despite of the above shortcomings, the $\QZcT$ still remains one of the best estimation of the FBW, since it can be measured, it is easy to evaluate, and provides good results in the majority of the cases. It is however recommended to antenna designers to be aware of the above mentioned problems, specifically in all the cases in which the measured / calculated $\QZcT$ yields smaller value than the $\QZcT$ of the pure \TM mode.}	
\end{itemize}


\section{Conclusion}
\label{Sec_Concl}

The potential theory has been employed to obtain the  quality factor $Q_Z$  of important spherical current distributions, particularly of the fundamental TM and TE modes and their combination. It has been shown that the presented approach is effective, leading to unique and finite energy terms with the far-field extraction implicitly included. For the presented cases of the spherical coordinate system,  the $Q_Z$  was obtained in closed form for any $ka$. The lower limit of  the $Q_Z$ of electrically small single-mode antennas was then obtained by series expansion of these expressions for small $ka$. Excellent agreement with the previous work of Thal and Hansen  has been observed. On contrary, the analysis of multimodal currents revealed that $Q_Z$ of the pure \TM mode cannot be considered as a true lower bound of $Q_Z$ of a general current radiating in free space. A particular example of linear combination of the \TM and the \TE mode has in fact showed that the $Q_Z$ can be tuned to values as low as $\QZcT \approx 1$ even for electrically very small structures.

The proposed approach has been presented on spherical modes, but it is not restricted to them, and can easily be extended to other separable coordinate systems. In this respect, the elliptic coordinates may be of considerable interest, as they can closely match the shape of many realistic antennas. The $Q_Z$ obtained in this way could then represent practically oriented limitations for antenna designers.

\appendices
\section{Vector and scalar potentials of the \TM and \TE mode}
\label{app2}

The vector and scalar potentials are found by the expansion method of appendix~\ref{app0}. For the particular case of the \TM and the \TE modes, we obtain the corresponding vector and scalar potentials regular for $r = [0,\infty)$ by using (\ref{EqExpan2_A})--(\ref{EqExpan2_C}) with $\MAT{a} = \MAT{z}_0$, $\psi_{00} = \MR{z}_0 (kr)$ for $\MAT{M}$-, $\MAT{N}$-terms and with $\psi_{10} = \MR{z}_1 (kr) \cos(\vartheta)$ for $\MAT{L}$-terms, where $\MR{z}_n(x)$ is a spherical Bessel function of order $n$ and where we will use $\MR{z}_n(x) = \MR{j}_n(x)$ for $r<a$ and $\MR{z}_n = \MR{h}_n^{(2)}$ for $r>a$. The resulting vector wave functions read
\begin{subequations}
\begin{align}
\label{Eq1A_app2}
\MAT{M}_{10} &= \boldsymbol\varphi_0 k \, \MR{z}_1 \left(kr \right) \sin\left(\vartheta \right), \\
\label{Eq1B_app2}
\MAT{N}_{10} &= \MAT{r}_0 \frac{2}{r} \MR{z}_1 \left(kr \right) \cos\left(\vartheta \right) \nonumber \\
&+ \boldsymbol\vartheta_0 \frac{1}{r} \left( \MR{z}_1 \left(kr \right) - kr\,\MR{z}_0 \left(kr \right) \right) \sin\left(\vartheta \right), \\
\label{Eq1C_app2}
\MAT{L}_{10} &= \MAT{r}_0 \frac{1}{r} \left( \MR{z}_1 \left(kr \right) - kr\,\MR{z}_2 \left(kr \right) \right) \cos\left(\vartheta \right) \nonumber \\
& - \boldsymbol\vartheta_0 \frac{1}{r} \MR{z}_1 \left(kr \right) \sin\left(\vartheta \right).
\end{align}
\end{subequations}
The vector potential of the \TM mode will be expressed as a linear combination of (\ref{Eq1B_app2}) and (\ref{Eq1C_app2}) because of the non-vanishing charge density and the need for the $\MAT{L}_{10}$-term. The vector potential of the \TE mode will be expressed in terms of (\ref{Eq1A_app2}) only, since there is no charge density and thus no scalar potential.

According to the above, in order to find the vector and the scalar potential of the \TM mode, we choose
\begin{subequations}
\begin{align}
\label{Eq2A_app2}
&\left. \begin{array}{ll} 
\MAT{A} &= C_1 \MAT{N}_{10} + D_1 \MAT{L}_{10},\\
\varphi &= -\J \omega D_1 \psi_{10}
\end{array} \quad \right\} r < a \\
& \nonumber \\
\label{Eq2B_app2}
&\left. \begin{array}{ll} 
\MAT{A} &= C_2 \MAT{N}_{10} + D_2 \MAT{L}_{10},\\
\varphi &= -\J \omega D_2 \psi_{10}
\end{array} \quad \right\} r > a,
\end{align}
\end{subequations}
where $C$, $D$ are constants to be determined. The $C_{1,2}$ can be determined from the boundary conditions on the current shell at $r = a$, i.e. by continuity of the tangential electric field \mbox{$\MAT{n}_0 \times \left( \MAT{E}_1 - \MAT{E}_2 \right) = 0$} and discontinuity of the tangential magnetic field \mbox{$\MAT{n}_0 \times \left( \MAT{H}_1 - \MAT{H}_2 \right) = \MAT{K}$}, where $\MAT{K}$ is the surface current density and where the normal $\MAT{n}_0$ points to the region 1, \cite{Jackson_ClassicalElectrodynamics}. The boundary conditions lead to
\begin{subequations}
\begin{align}
\label{Eq3A_app2}
C_1 &= -\frac{\J \mu}{2 \pi k} \left( \MR{h}_1^{(2)} \left(ka \right) - ka \, \MR{h}_0^{(2)} \left(ka \right) \right), \\
\label{Eq3B_app2}
C_2 &= -\frac{\J \mu}{2 \pi k} \left( \MR{j}_1 \left(ka \right) - ka \, \MR{j}_0 \left(ka \right) \right).
\end{align}
\end{subequations}
For the unknown constants $D_{1,2}$ in (\ref{Eq2A_app2}) and (\ref{Eq2B_app2}), the only condition that needs to be satisfied is the wave equation for the scalar potential in the Lorentz gauge
\BE
\label{Eq4_app2}
\nabla^2 \varphi + k^2 \varphi = - \frac{\rho}{\epsilon}.
\EE
Choosing then the scalar potential being continuous at $r = a$, (\ref{Eq4_app2}) dictates
\BE
\label{Eq5_app2}
\frac{\partial \varphi}{\partial r}\Big\vert_{r=a^+} - \frac{\partial \varphi}{\partial r}\Big\vert_{r=a^-} = - \J \frac{\cos\left(\vartheta\right)}{\omega\epsilon\pi a^2},
\EE
which leads to
\begin{subequations}
\begin{align}
\label{Eq6A_app2}
D_1 &= \frac{\J \mu}{k \pi} \MR{h}_1^{(2)} \left( ka \right), \\
\label{Eq6B_app2}
D_2 &= \frac{\J \mu}{k \pi} \MR{j}_1 \left( ka \right).
\end{align}
\end{subequations}
Putting all together we have for $r < a$
\begin{subequations}
\begin{align}
\label{Eq7A_app2}
A_\vartheta &= -\frac{\J \mu}{2 \pi kr} \sin \left( \vartheta \right) \Bigg( 2 \MR{h}_1^{(2)} \left( ka\right) \MR{j}_1 \left( kr\right) \\
&+ \bigg( \MR{h}_1^{(2)} \left( ka\right) - ka\, \MR{h}_0^{(2)} \left( ka\right) \bigg) \bigg( \MR{j}_1 \left( kr\right) - kr \, \MR{j}_0 \left( kr\right) \bigg) \Bigg) \nonumber \\
\label{Eq7B_app2}
\varphi &= \frac{\omega\mu}{\pi k} \MR{h}_1^{(2)} \left( ka\right) \MR{j}_1 \left( kr\right) \cos \left( \vartheta \right),
\end{align}
\end{subequations}
and for $r > a$
\begin{subequations}
\begin{align}
\label{Eq8A_app2}
A_\vartheta &= -\frac{\J \mu}{2 \pi kr} \sin \left( \vartheta \right) \Bigg( 2 \MR{h}_1^{(2)} \left( kr\right) \MR{j}_1 \left( ka\right) \\
&+ \bigg( \MR{h}_1^{(2)} \left( kr\right) - kr\, \MR{h}_0^{(2)} \left( kr\right) \bigg) \bigg( \MR{j}_1 \left( ka\right) - ka \, \MR{j}_0 \left( ka\right) \bigg) \Bigg) \nonumber \\
\label{Eq8B_app2}
\varphi &= \frac{\omega\mu}{\pi k} \MR{h}_1^{(2)} \left( kr\right) \MR{j}_1 \left( ka\right) \cos \left( \vartheta \right),
\end{align}
\end{subequations}
where the first terms in the vector potential come from $\MAT{L}_{10}$ and the second terms come from $\MAT{N}_{10}$.

The derivation of the scalar and vector potential of the \TE mode is analogous to the above, and results in
\begin{subequations}
\begin{align}
\label{Eq9A_app2}
A_\varphi &= -\frac{\J \mu}{2} \sin\left(\vartheta\right) ka \, \MR{j}_1 \left( kr \right) \MR{h}_1^{(2)} \left( ka \right) \\
\label{Eq9B_app2}
\varphi &= 0,
\end{align}
\end{subequations}
for $r < a$, and
\begin{subequations}
\begin{align}
\label{Eq10A_app2}
A_\varphi &= -\frac{\J \mu}{2} \sin\left(\vartheta\right) ka \, \MR{j}_1 \left( ka \right) \MR{h}_1^{(2)} \left( kr \right) \\
\label{Eq10B_app2}
\varphi &= 0,
\end{align}
\end{subequations}
for $r < a$.

\section{Expansion of the vector wave equation in separable systems}
\label{app0}

In this appendix, we recall the expansion of the vector wave equation and point out some aspects important for a consistent definition of Q. This approach leads to the analytical calculation of the Q factor for the separable systems. 

According to \cite{Hansen_NewTypeOfExpanstion}, the general solution of \mbox{$\nabla^2 \MAT{A} + k^2 \MAT{A} = 0$} can be written as
\BE
\label{EqExpan1}
\MAT{A} = \sum\limits_n \left( \alpha_n \MAT{M}_n + \beta_n \MAT{N}_n + \gamma_n \MAT{L}_n \right),
\EE
where
\begin{subequations}
\begin{align}
\label{EqExpan2_A}
\MAT{M}_n &= \nabla\times\left( \MAT{a} \psi_n \right), \\
\label{EqExpan2_B}
\MAT{N}_n &= \frac{1}{k} \nabla\times \MAT{M}_n, \\
\label{EqExpan2_C}
\MAT{L}_n &= \nabla \psi_n,
\end{align}
\end{subequations}
$\MAT{a}$ is a constant vector, and scalar function $\psi_n$ satisfies
\BE
\label{EqExpan3}
\nabla^2 \psi_n + k^2 \psi_n = 0.
\EE
The conventional notation from \cite{Stratton_ElectromagneticTheory} is used for clarity of the paper.

Taking now the vector field $\MAT{A}$ as a magnetic vector potential in the Lorentz gauge, one can verify that the scalar potential is
\BE
\label{EqExpan4}
\varphi = - \frac{1}{\J \omega \epsilon \mu} \nabla \cdot \MAT{A} = - \J \omega \sum\limits_n \gamma_n \psi_n,
\EE
where $\mu$ is the permeability of the vacuum, and that the field quantities read
\begin{subequations}
\begin{align}
\label{EqExpan5_A}
\MAT{E} &= -\frac{\J \omega}{k^2} \left( \nabla \nabla \cdot \MAT{A} + k^2 \MAT{A} \right) = - \J \omega \sum\limits_n \left( \alpha_n \MAT{M}_n + \beta_n \MAT{N}_n \right), \\
\label{EqExpan5_B}
\MAT{H} &= \frac{1}{\mu} \nabla\times \MAT{A} = \frac{k}{\mu} \sum\limits_n \left( \alpha_n \MAT{N}_n + \beta_n \MAT{M}_n \right),
\end{align}
\end{subequations}
where we used the fact that \mbox{$\nabla\times\MAT{N}_m = k \MAT{M}_n$}.

It is worth noting that any measurable quantity is independent of $\MAT{L}_n$-terms (which is equivalent to gauge invariance). Particularly, if a volume is chosen so that it contains all the sources and if the vector potential (\ref{EqExpan1}) is divided as $\MAT{A} = \MAT{A}^{M,N} + \MAT{A}^{L}$, with $\MAT{A}^{M,N}$ belonging to $\MAT{M}_n$-, $\MAT{N}_n$-terms and $\MAT{A}^{L}$ belonging to $\MAT{L}_n$-terms, then, one can easily realize that
\BE
\label{EqExpan6}
\int\limits_\Omega \left( \MAT{A}^L \cdot \MAT{J}^\ast - \varphi \rho^\ast \right) \D{\mathbf{r}} = 0,
\EE
and thus that only the $\MAT{M}_n$-, $\MAT{N}_n$-terms participate in the definition of the complex power \cite{Jackson_ClassicalElectrodynamics}
\BE
\label{EqExpan7}
\J \omega \int\limits_\Omega \left( \MAT{A} \cdot \MAT{J}^\ast - \varphi \rho^\ast \right) \D{\mathbf{r}} = \J \omega \int\limits_\Omega \left( \MAT{A}^{M,N} \cdot \MAT{J}^\ast \right) \D{\mathbf{r}}.
\EE


\section{Analytical Expressions for the $\QZcT$ and the $\Qhansen$ of the \TM and the \TE Modes}
\label{app3}

This appendix presents the analytical expressions for the $\Qhansen$ and the $\QZcT$ for the dominant spherical \TM and \TE modes. The $\QZcT$ is evaluated according to the method introduced in this paper (Sections~III,~IV,~V). The $\Qhansen$ is evaluated using the classical extraction method of Collin and Rothschild \cite{CollinRotchild_EvaluationOfAntennaQ} (Hansen and Collin \cite{HansenCollin_ANewChuFormulaForQ}).

Particularly, the expressions for the \TM mode read
\begin{equation}
\label{TM10extr_Eq1}
\Qhansen = \max \displaystyle\left\{ \begin{array}{c}
\displaystyle\frac{2 \left( x^5 - x^3 + x \right) - 4 x \left( x^2 - 2 \right) \Cx{2} - \left(x^4 - 9 x^2 + 5\right) \Sx{2}}{4 \left(x \Cx{} + \left( x^2 -1 \right) \Sx{}\right)^2},\\
\\
\displaystyle\frac{2 \left( x^5 - x^3 + x \right) + 4 x \Cx{2} + \left(x^4 + 3 x^2 -3 \right) \Sx{2}}{4 \left(x \Cx{} + \left( x^2 -1 \right) \Sx{}\right)^2}
\end{array} \right\},
\end{equation}
\\
\begin{equation}
\label{TM10extr_Eq2}
\begin{array}{ll}
\QZcT &= \displaystyle\frac{{\left| {x^5 - 2 \J x^4 - 4 x^3 + 5 \J x^2 + 4 x - 2 \J + \mathrm{e}^{2 \J x} \left( {2 \J - \J x^2 + \cdots } \right.} \right.}}{2 \left( x \Cx{} + \cdots \right.} \\
\\
&\quad\displaystyle\frac{{\left. {\left. { \ldots  + \left| \left(x \Sx{} - \left(x^2 - 1\right) \Cx{}\right) \left( x \Cx{} + \left(x^2 - 1\right)\Sx{}\right) \right|} \right)} \right|}}{{\left. \cdots + \left( x^2 - 1 \right) \Sx{} \right)^2}},
\end{array}
\end{equation}
while the expressions for the \TE mode read
\begin{equation}
\label{TE10extr_Eq1}
\Qhansen = \max\left\{ \frac{2 x \left( x^2 + 1 \right) + 4 x \Cx{2} + \left( x^2 - 3 \right) \Sx{2}}{4 \left(\Sx{} - x \Cx{} \right)^2}, \frac{\left(x^2 + 1 \right) \left( 2 x - \Sx{2}\right)}{4 \left(\Sx{} - x \Cx{} \right)^2} \right\},
\end{equation}
\\
\begin{equation}
\label{TE10extr_Eq2}
\QZcT = \frac{\left| x^3 - 2 \J x^2 - 2 x + \J - \mathrm{e}^{2 \J x} \left( \J + \left| \left( x \Cx{} - \Sx{} \right) \left( x \Sx{} + \Cx{} \right) \right| \right) \right|}{2 \left(\Sx{} - x \Cx{} \right)^2}.
\end{equation}

\section*{Acknowledgement}
This work has been supported by the Czech Science Foundation under projects No. P102/12/2223 and \mbox{13-09086S}. The authors would like to thank to Mats Gustafsson from Lund University, Sweden for pointing out the possibility of $Q_Z$ minimization by the combination of modes.

\ifCLASSOPTIONcaptionsoff
  \newpage
\fi

\end{document}